\documentclass[aps,prd,twocolumn,superscriptaddress]{revtex4-1}

\usepackage{amssymb}
\usepackage{amsmath}
\usepackage{graphicx}
\usepackage{enumerate}
\usepackage{mathtools}
\usepackage{color}
\usepackage{bbold}
\usepackage{subfigure}
\usepackage{slashed}
\usepackage{xcolor}
\usepackage{bm}
\usepackage{ulem}

\begin{document}


\title{Neutrino oscillations in supernovae: angular moments and fast instabilities}


\author{Lucas Johns}
\email[]{ljohns@physics.ucsd.edu}
%\thanks{}
\affiliation{Department of Physics, University of California, San Diego, La Jolla, California 92093, USA}

\author{Hiroki Nagakura}
\affiliation{Department of Astrophysical Sciences, Princeton University, 4 Ivy Lane, Princeton, NJ 08544, USA}

\author{George M. Fuller}
\affiliation{Department of Physics, University of California, San Diego, La Jolla, California 92093, USA}

\author{Adam Burrows}
\affiliation{Department of Astrophysical Sciences, Princeton University, 4 Ivy Lane, Princeton, NJ 08544, USA}


%\date{\today}

\begin{abstract}
Recent theoretical work indicates that the neutrino radiation in core-collapse supernovae may be susceptible to flavor instabilities that set in far behind the shock, grow extremely rapidly, and have the potential to profoundly affect supernova dynamics and composition. Here we analyze the nonlinear collective oscillations that are prefigured by these instabilities. We demonstrate that a zero-crossing in $n_{\nu_e} - n_{\bar{\nu}_e}$ as a function of propagation angle is not sufficient to generate instability. Our analysis accounts for this fact and allows us to formulate complementary criteria. Using \textsc{Fornax} simulation data, we show that fast collective oscillations qualitatively depend on how forward-peaked the neutrino angular distributions are.
\end{abstract}

%\pacs{ab.cd, 12.34}
%\keywords{}

\maketitle

%\section{}
%\subsection{}
%\subsubsection{}

%\section{Section}

In this paper we address a key aspect of neutrino physics in core-collapse supernovae. The stakes are high, as supernova explosions are central to our understanding of the origin of elements and the history of galaxies.

Recently it has been realized that the neutrino flavor field in core-collapse supernovae is prone to a host of instabilities \cite{cherry2012, raffelt2013, chakraborty2014, mangano2014, duan2015b, abbar2015, abbar2015b, dasgupta2015, chakraborty2016, capozzi2017, izaguirre2017} that were artificially concealed by the symmetries adopted in older studies \cite{duan2006, hannestad2006, fogli2007, esteban2008, dasgupta2009, duan2010, dasgupta2012}. Of particular urgency is the subclass known as \textit{fast} instabilities, so named because they exhibit growth rates proportional to the self-coupling potential $\mu = \sqrt{2} G_F n_\nu$ and are not suppressed by the typically much smaller vacuum oscillation frequency $\omega = \delta m^2 / 2 E$ \cite{sawyer2005, sawyer2009, sawyer2016, chakraborty2016b, chakraborty2016c, dasgupta2017, dasgupta2018, dasgupta2018b, airen2018, abbar2018, abbar2019, capozzi2019, capozzi2019b, yi2019, martin2019, chakraborty2019}. They are commonly, if not always, associated with zero-crossings of the electron lepton number carried by neutrinos ($\nu$ELN) as a function of propagation angle. Global variations in $n_{\bar{\nu}_e} / n_{\nu_e}$---possibly related to LESA (lepton-number emission self-sustained asymmetry) \cite{tamborra2014, walk2018, oconnor2018, vartanyan2018, glas2019, vartanyan2019, walk2019}---and coherent neutrino--nucleus scattering \cite{morinaga2019} independently make this condition a live possibility in core-collapse supernovae \cite{tamborra2017, abbar2019b, azari2019, shalgar2019, morinaga2019, nagakura2019}. If fast flavor conversion (FFC) does occur, it could substantially alter our current view of supernova dynamics and nucleosynthesis \cite{fuller1987, fuller1992}.

The aim of the present study is to gain some degree of understanding of the nonlinear collective effects heralded by fast flavor instabilities. Our basic approach is to study the evolution of the neutrino flavor field in terms of its momentum-space angular moments. Three considerations motivate this choice. The first is realism: Neutrino angular distributions within $\sim 100$ km are quite unlike the forms they are given in bulb or beam models. In point of fact, they transition---very gradually relative to the $\mu^{-1}$ scale---from nearly isotropic to narrowly forward-peaked \cite{thompson2003, ott2008, nagakura2018, harada2019}. The second consideration is computational: As it is, many state-of-the-art supernova simulations only track the first few angular moments, and cohesion between hydrodynamic and oscillation calculations is clearly desirable \cite{strack2005, zhang2013, notzold1988, volpe2013, vlasenko2014, kartavtsev2015, richers2019, stapleford2019}. The last is theoretical: In multipole space, the factor $\left( 1 - \mathbf{\hat{p}}\cdot \mathbf{\hat{q}} \right)$ that couples neutrinos of momenta $\mathbf{p}$ and $\mathbf{q}$ becomes a sum of monopole and dipole couplings \cite{raffelt2007b, duan2014}. Angular moments are consequently a natural lens through which to examine collective oscillations.

This last observation is especially true of fast modes, which can be isolated by taking $\mu \gg \omega$. Because neutrino energy drops out of the coherent evolution, we can work with polarization vectors that are integrated over the spectrum. Neutrinos propagating in a homogeneous environment at angle $v = \cos \theta$ (axial symmetry is assumed throughout) then obey the hybrid multipole/momentum equation
\begin{equation}
\dot{\mathbf{P}}_v = \mu \left( \mathbf{D}_0 - v \mathbf{D}_1 \right) \times \mathbf{P}_v. \label{hybrid}
\end{equation}
Here $\mathbf{D}_0$ and $\mathbf{D}_1$ are the monopole and dipole difference vectors ($\mathbf{D}_l = \mathbf{P}_l - \mathbf{\bar{P}}_l$) and  the matter potential $\lambda = \sqrt{2} G_F n_e$ has been rotated out. It is immediately apparent that the only way for the flavor content $P_{v,z}$ to change significantly on a fast time scale is for $\mathbf{D}_0 - v \mathbf{D}_1$ to swing away from the flavor axis.

\begin{figure}
\centering
\begin{subfigure}{
\centering
\includegraphics[width=.46\textwidth]{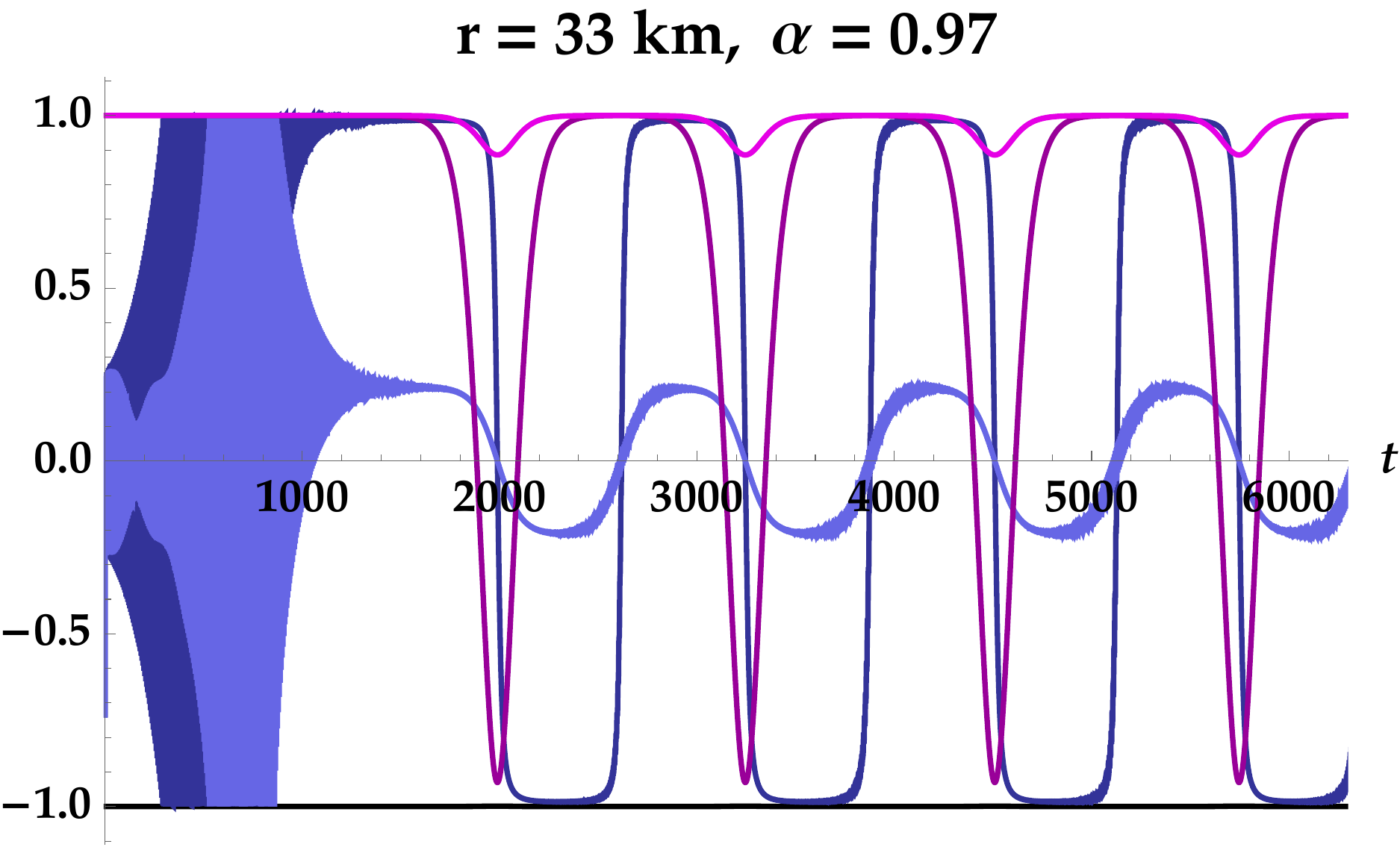}
}
\end{subfigure}
\begin{subfigure}{
\centering
\includegraphics[width=.46\textwidth]{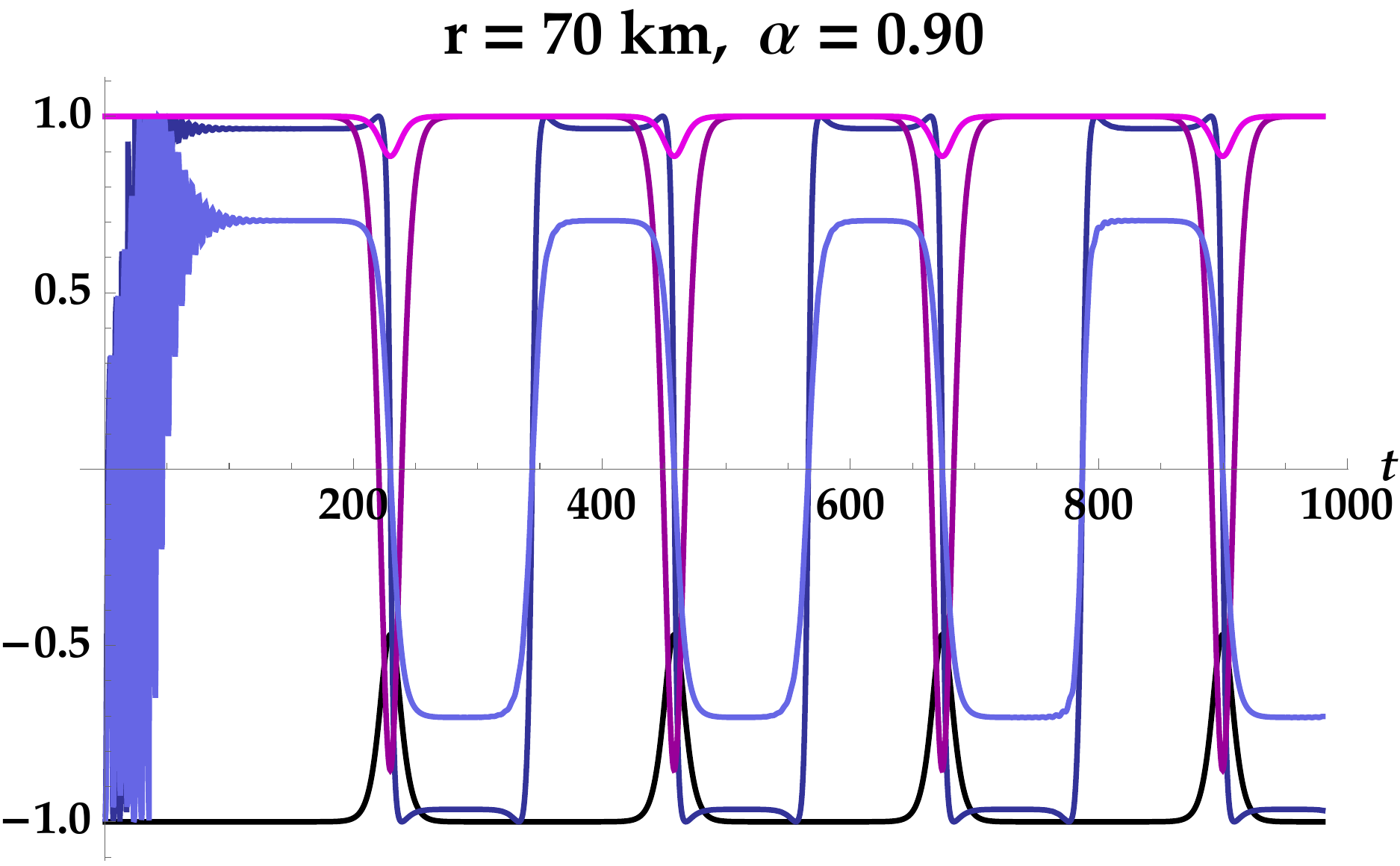}
}
\end{subfigure}
\caption{Angular coordinates over four periods of fast flavor conversion. Two values of $v = \cos \theta$ are shown in each panel. The one that experiences more significant flavor conversion is distinguished by the use of darker shades: purple for $\cos\theta_v$, blue for $\sin \left( \phi_v - \phi_1 \right)$. The thick black curve depicts $\cos\theta_1$. Time is in units of $[ \sqrt{2} G_F ( n_{\nu_e} - n_{\bar{\nu}_e} )]^{-1} \sim 14$ ps (154 ps) for the upper (lower) panel. See the text for discussion and Fig.~\ref{flavor_time} for more information on the choice of parameters.}  
\label{angles}
\end{figure}

\begin{figure*}
\centering
\begin{subfigure}{
\centering
\includegraphics[height=1.8 in]{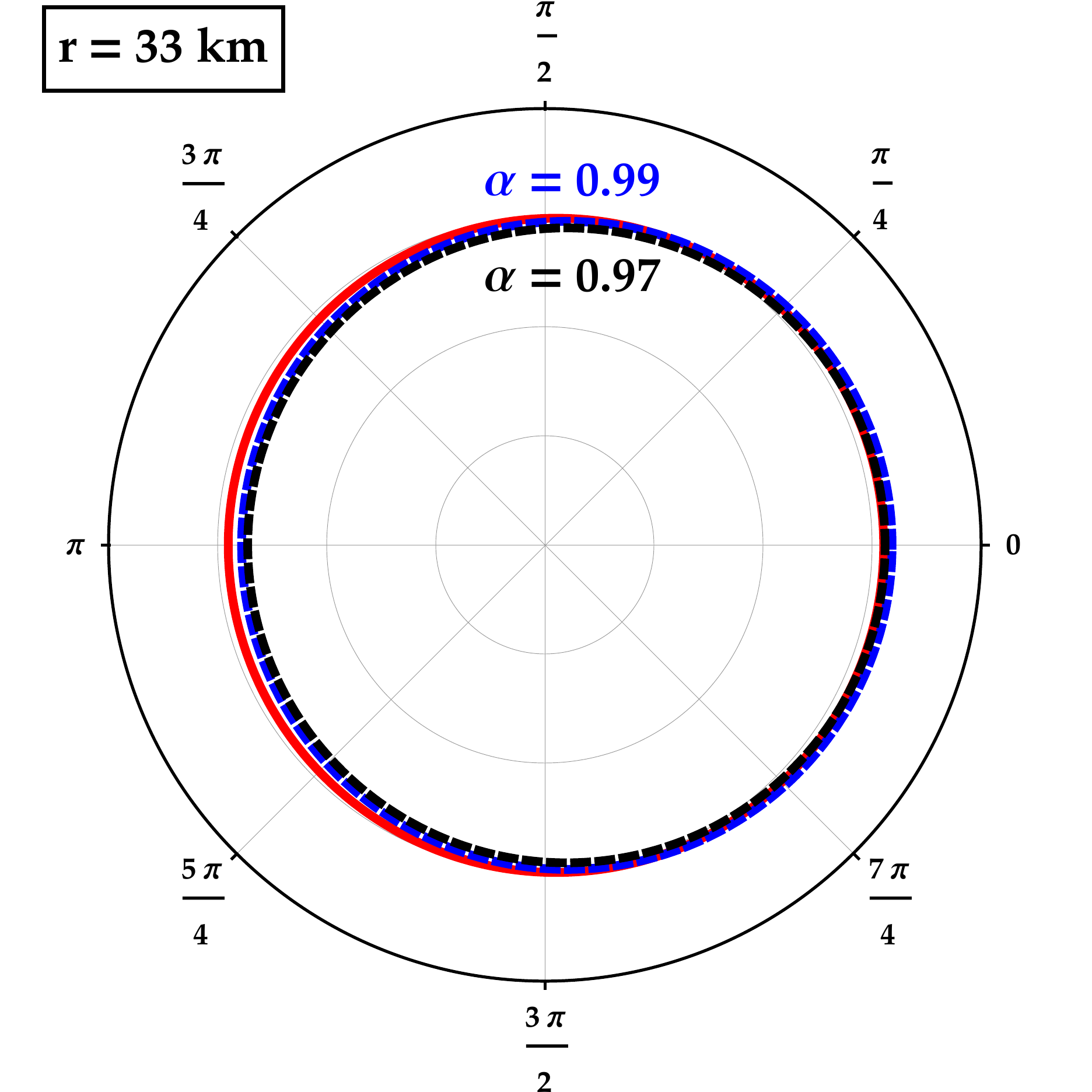}
}
\end{subfigure}
\begin{subfigure}{
\centering
\includegraphics[height=1.8 in]{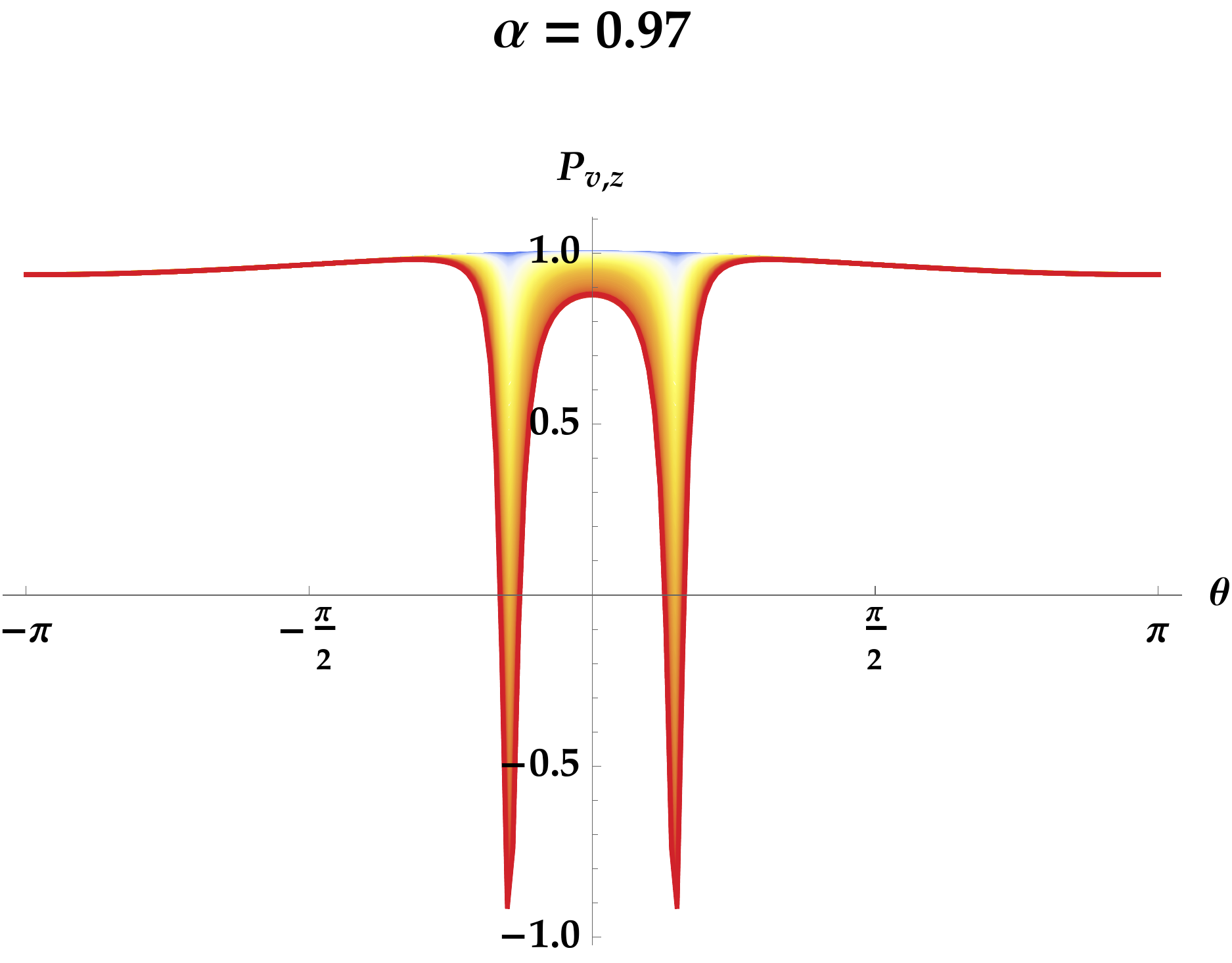}
}
\end{subfigure}
\begin{subfigure}{
\centering
\includegraphics[height = 1.8 in]{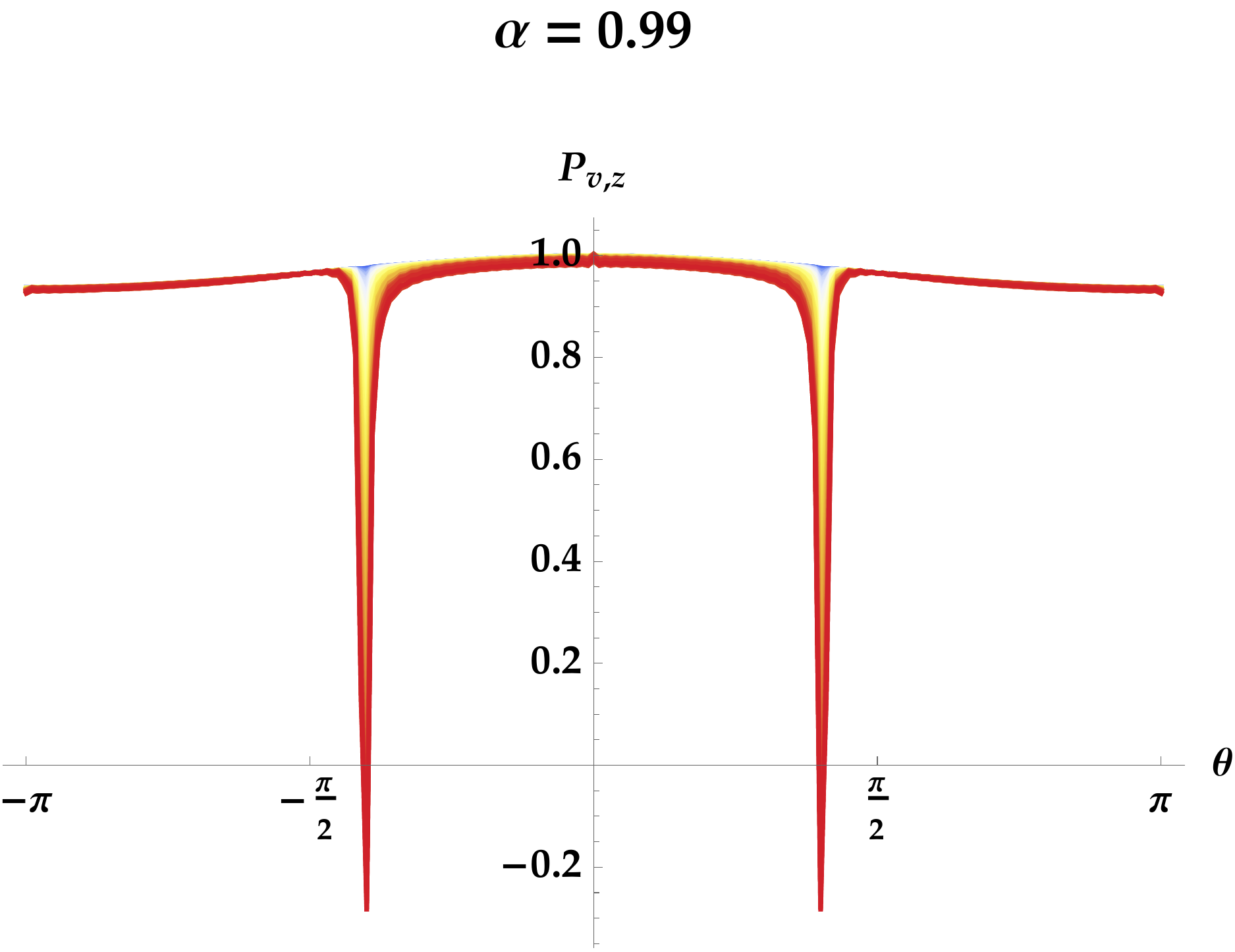}
}
\end{subfigure}

\begin{subfigure}{
\centering
\includegraphics[height=1.8 in]{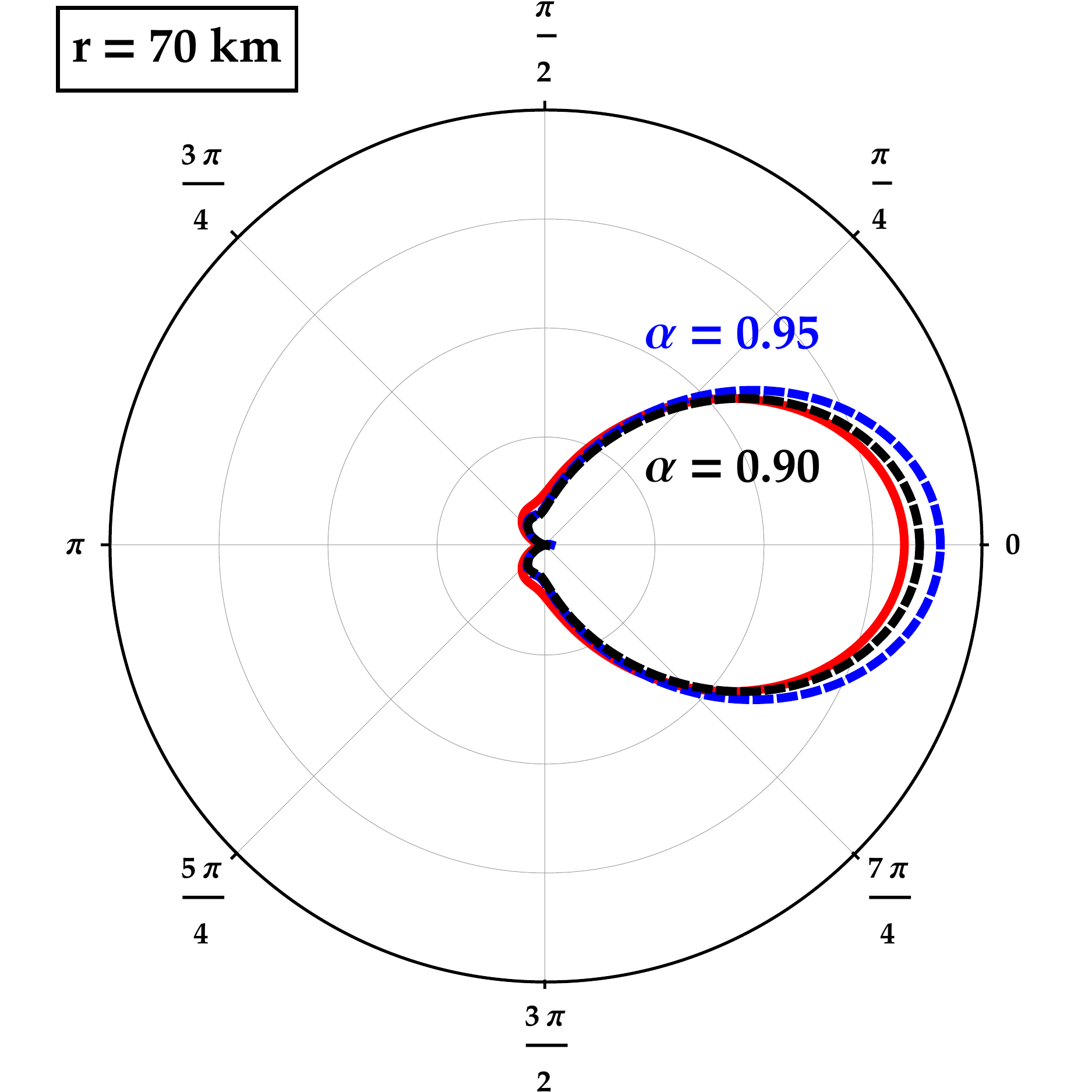}
}
\end{subfigure}
\begin{subfigure}{
\centering
\includegraphics[height = 1.8 in]{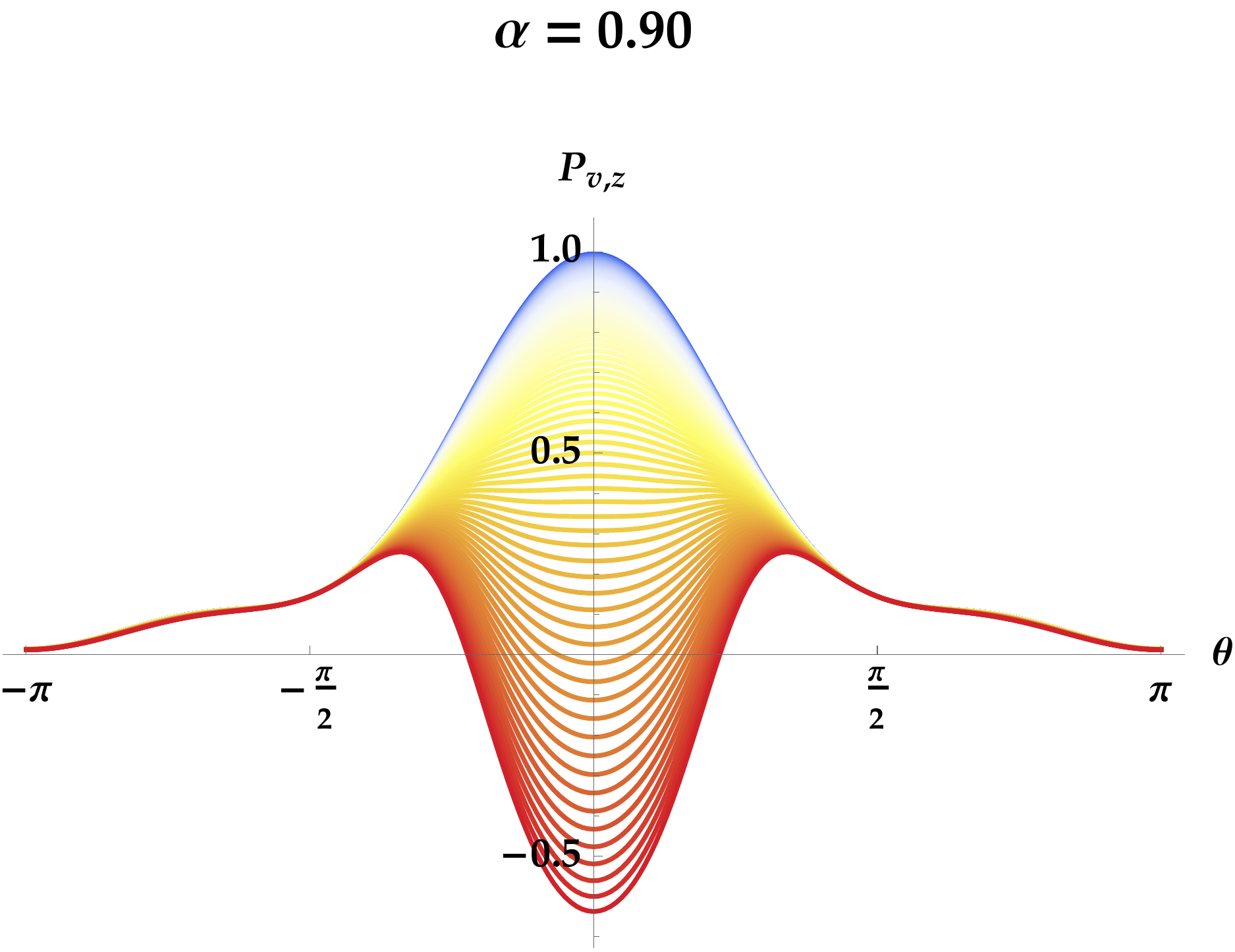}
}
\end{subfigure}
\begin{subfigure}{
\centering
\includegraphics[height = 1.8 in]{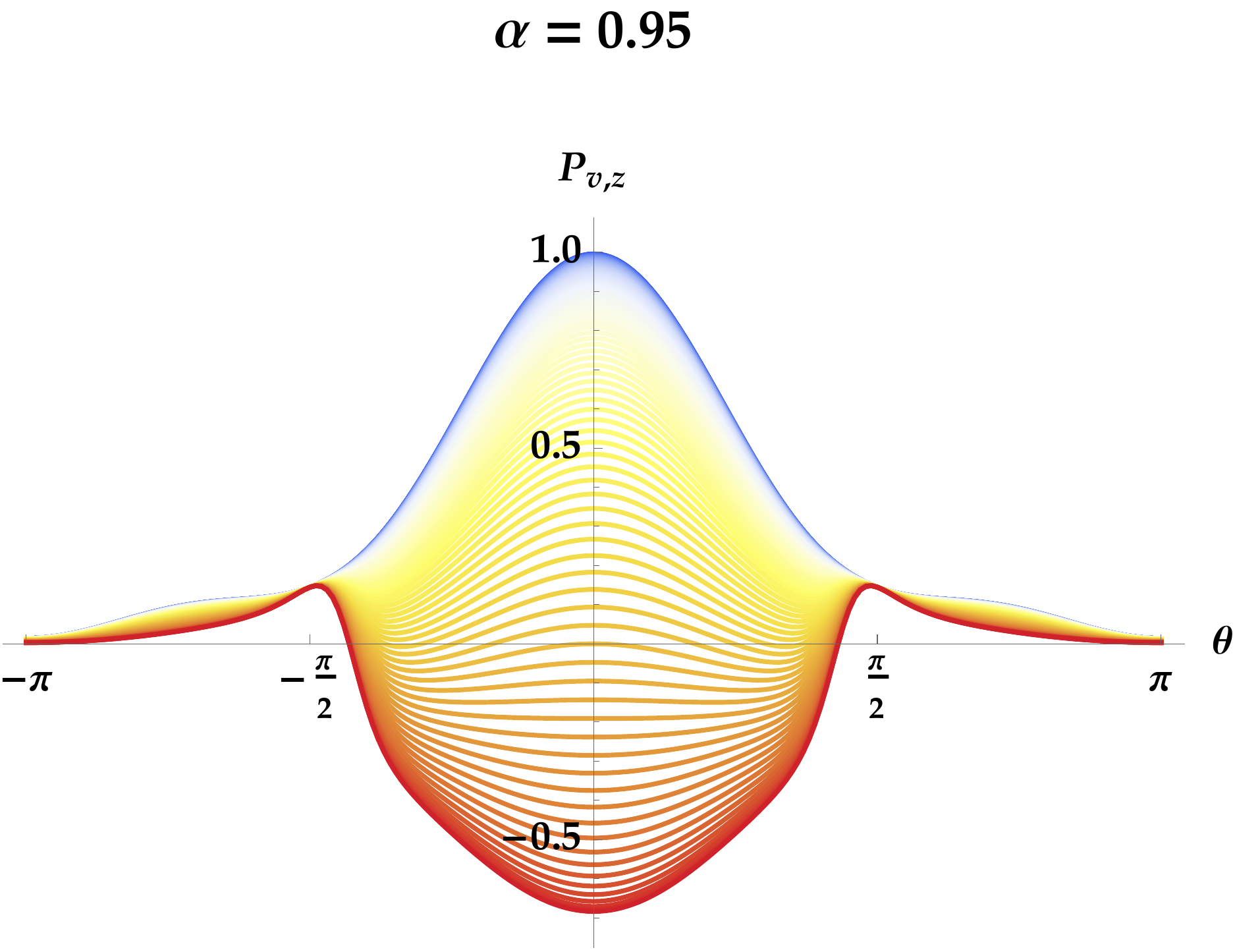}
}
\end{subfigure}
\caption{\textit{Left}: $n_{\nu_e}$ (red) and $n_{\bar{\nu}_e}$ (blue and black) as functions of propagation angle $\theta$, with arbitrary normalization. The angular distributions are drawn at 200 ms post-bounce from a spherically symmetric \textsc{Fornax} simulation \cite{radice2017, skinner2019} of the $16~M_\odot$ progenitor from Ref.~\cite{sukhbold2016}. M1 closure is used to provide the radiative pressures and radiative heat fluxes \cite{vaytet2011}, and $\alpha = n_{\bar{\nu}_e} / n_{\nu_e}$ is treated as a free parameter in order to trigger instability. \textit{Middle and right}: Snapshots of $P_{v,z}$ color-coded by time (going from blue to red) and spanning the descent phase of a single dip in $D_{1,z}$. The normalization is such that $P_{v=1, z} = 1$. To isolate the fast mode, $\omega$ is assigned an artificially small value.}
\label{flavor_time}
\end{figure*}

In terms of the difference vectors and their counterpart sum vectors $\mathbf{S}_l = \mathbf{P}_l + \mathbf{\bar{P}}_l$, the multipole equations of motion are \cite{raffelt2007b}
\begin{align}
&\dot{\mathbf{S}}_l = \mu \mathbf{D}_0 \times \mathbf{S}_l - \frac{\mu}{2} \mathbf{D}_1 \times \left( a_l \mathbf{S}_{l-1} + b_l \mathbf{S}_{l+1} \right), \notag \\
&\dot{\mathbf{D}}_l = \mu \mathbf{D}_0 \times \mathbf{D}_l - \frac{\mu}{2} \mathbf{D}_1 \times \left( a_l \mathbf{D}_{l-1} + b_l \mathbf{D}_{l+1} \right), \label{sumdiff}
\end{align}
where $a_l = 2l / (2l+1)$ and $b_l = 2 (l+1) / (2l+1)$. $\mathbf{D}_0$ is constant on $\mu^{-1}$ time scales, implying that fast collective modes must be driven by $\mathbf{D}_1$. It is helpful at this point to switch to a frame rotating about $\mathbf{\hat{D}}_0$ at frequency $\mu D_0$, where $D_0 = | \mathbf{D}_0 |$. Using primes to denote vectors in the rotating frame and introducing $\mathbf{L}' = ( \mathbf{D}'_0 + 2 \mathbf{D}'_2 ) / 3$ and $\mathbf{G}' = 2 \mathbf{D}'_3 / 5$, we then have
\begin{align}
&\dot{\mathbf{D}}'_1 = \mu \mathbf{L}' \times \mathbf{D}'_1, \notag \\
&\dot{\mathbf{D}}'_2 = \frac{3}{2} \mu \mathbf{G}' \times \mathbf{D}'_1.
\end{align}
Computing $\mathbf{D}'_1 \times \ddot{\mathbf{D}}'_1$ leads to a pendulum equation, which can be written in a form comparable to that of the bipolar pendulum (Eq.~39 of Ref.~\cite{hannestad2006}) by defining $\boldsymbol{\delta}' = \mathbf{D}'_1 / D_1$ and $\sigma = \boldsymbol{\delta}' \cdot \mathbf{L}'$.
The result is
\begin{equation}
\frac{\boldsymbol{\delta}' \times \ddot{\boldsymbol{\delta}}'}{\mu} + \sigma \dot{\boldsymbol{\delta}}' = \mu D_1 \boldsymbol{G}' \times \boldsymbol{\delta}'. \label{deltaeq}
\end{equation}
One critical distinction with respect to the bipolar pendulum is that in this case ``gravity'' is not a fixed external potential. In fact, $\mathbf{G}'$ couples directly to $\mathbf{D}'_1$, making this a sort of \textit{nonlinear} gyroscopic pendulum. Nevertheless, the possibility for collective pendulum motion is built into the structure of Eq.~\eqref{sumdiff}. Numerical realizations of it are shown in Figs.~\ref{angles} and \ref{flavor_time}.

The dynamics of the system is also restricted by a tower of conservation laws, which can be constructed by differentiating $\mathbf{D}'_1 \cdot \mathbf{D}'_l$ and recursively reducing the right-hand side until it is expressed as a total derivative. The first three conserved quantities are $D_1$, $\sigma$, and
\begin{equation}
E_D = \mu  \boldsymbol{G}' \cdot \mathbf{D}_1' + \frac{\mu}{2} \mathbf{L}'^2, \label{conserved}
\end{equation}
which respectively denote the length of the pendulum, its spin, and its total energy. In a foundational study, Raffelt and Sigl \cite{raffelt2007b} showed that the dipole term is the driving force behind kinematic decoherence. This remains true on short time scales, and it is clear from Eq.~\eqref{hybrid} that $\mathbf{D}_1$ causes dephasing of neutrinos with different values of $v$. But the constraints on the motion of $\mathbf{D}_1$ mean that the dephasing can give rise to persistent collective oscillations rather than effectively irreversible relaxation, at least until the effects of finite $\omega$ become important. The additional fact that some of these constraints involve only the first four angular moments gives us some hope of capturing the important features of FFC without having fine-grained information about the distributions in momentum space. Indeed, the higher conservation laws, which encode the fact that all angular moments are dynamically linked, may have utility for closing the moment hierarchy in a sensible way.

We can be more specific about the connection to kinematic decoherence by recalling that $\mathbf{S}_0$ obeys a pendulum equation as well \cite{hannestad2006, duan2007b, raffelt2007b, johns2018}, with energy
\begin{equation}
E_S = \omega \mathbf{B} \cdot \mathbf{S}_0 + \frac{\mu}{2} \left( \mathbf{D}_0^2 - \mathbf{D}_1^2 \right).
\end{equation}
Kinematic decoherence arises because $\mathbf{D}_0^2$ and $\mathbf{D}_1^2$ are able to evolve at the cost of $\mathbf{S}_0$ shrinking \cite{raffelt2007b}. But if $\mu \gg \omega$, then the $\mathbf{S}_0$ pendulum generally has very little sway over the $\mathbf{D}_1$ pendulum. The opposite is not true, however: $\mathbf{D}_1$ steers the evolution of $\mathbf{S}_0$. Relaxation occurs through the mutual interaction of the two pendula; the fact that the influence is one-way in the $\omega \rightarrow 0$ limit enables sustained collective motion.

It remains for us to understand how the predilection of $\mathbf{D}'_1$ for pendulum motion is expressed through the individual polarization vectors. Ultimately our interest is in the projection onto the flavor axis:
\begin{equation}
\dot{P}_{v,z} = - \mu v \left( \mathbf{D}_1 \times \mathbf{P}_v \right)_z. \label{pvz}
\end{equation}
Writing the vectors in terms of their angular coordinates ($\theta_v$ and $\phi_v$ being the polar and azimuthal angles of $\mathbf{P}_v$, $\theta_1$ and $\phi_1$ being the same of $\mathbf{D}_1$), Eq.~\eqref{pvz} becomes
\begin{equation}
\dot{\theta}_v = \mu v D_1 \sin \theta_1 \sin \left( \phi_v - \phi_1 \right). \label{thetav}
\end{equation}
Approximating $\phi_v$ and $\phi_1$ as developing under the influence of their \textit{initial} Hamiltonians, the phase difference accumulates at a rate
\begin{equation}
\dot{\phi}_v - \dot{\phi}_1 \simeq - \mu \left( \frac{1}{3} D_{0,z} (0) + v D_{1,z} (0) + \frac{2}{3} D_{2,z} (0) \right). \label{phasediff}
\end{equation}
Suppose that $\mathbf{P}_v (0) \propto \mathbf{z}$. If the phase difference develops slowly enough that the right-hand side of Eq.~\eqref{thetav} is positive over many cycles of $\phi_1$, then $\theta_v$ can grow to a size unsuppressed by the vacuum mixing angle.

As the instability grows, Eq.~\eqref{phasediff} breaks down and is replaced by the collective motion seen in Fig.~\ref{angles}. $P_{v,z}$ dips in proportion to $v \sin (\phi_v - \phi_1)$ and is reflected in---and driven by---peaks in $D_{1,z}$ (which are imperceptibly small in the upper panel because the angular distributions are very nearly isotropic). As Fig.~\ref{flavor_time} illustrates, there are two qualitatively different outcomes as a function of $v$. Setting Eq.~\eqref{phasediff} equal to zero, we find the trajectory which in this approximation has constant phase with respect to $\mathbf{D}_1$:
\begin{equation}
\tilde{v} = -\frac{1}{3 R_1} - \frac{2 R_2}{3 R_1}, \label{vtilde}
\end{equation}
with $R_{l} = D_{l,z} (0) / D_{0,z} (0)$. The quantity $\tilde{v}$ serves as a control parameter that shapes the $v$-dependence of the collective oscillations. When $\tilde{v}$ is comfortably inside the range $[-1,1]$ (as in the test cases at $r = 33$ km), it indicates the presence of narrow resonances. When $\tilde{v} \approx \pm 1$ (as at $r = 70$ km), the resonances fuse. Going one step further, we can use this parameter as the basis for a simple stability criterion: If $|\tilde{v}| > 1$, FFC cannot occur.

%Motivated by the equations of motion, let us make the following ansatz:
%\begin{align}
%&\phi_v (t) = \phi_v (0) + \mu \left[ D_{0,z} (0) - v D_{1,z} (0) \right] t, \notag \\
%&\phi_1 (t) = \phi_1 (0) + \frac{2}{3} \mu \left[ 2 D_{0,z} (0) + D_{2,z} (0) \right] t.
%\end{align}
%Ignoring an initial phase difference, it follows that
%\begin{align}
%\dot{\theta}_v = &- \mu v D_1 \sin \theta_1 \notag \\
%&\times \sin \left[ \mu \left( \frac{1}{3} D_{0,z} (0) + v D_{1,z} (0) + \frac{2}{3} D_{2,z} (0) \right) t \right]. \label{thetav}
%\end{align}

\begin{figure*}
\centering
\begin{subfigure}{
\centering
\includegraphics[width=.32\textwidth]{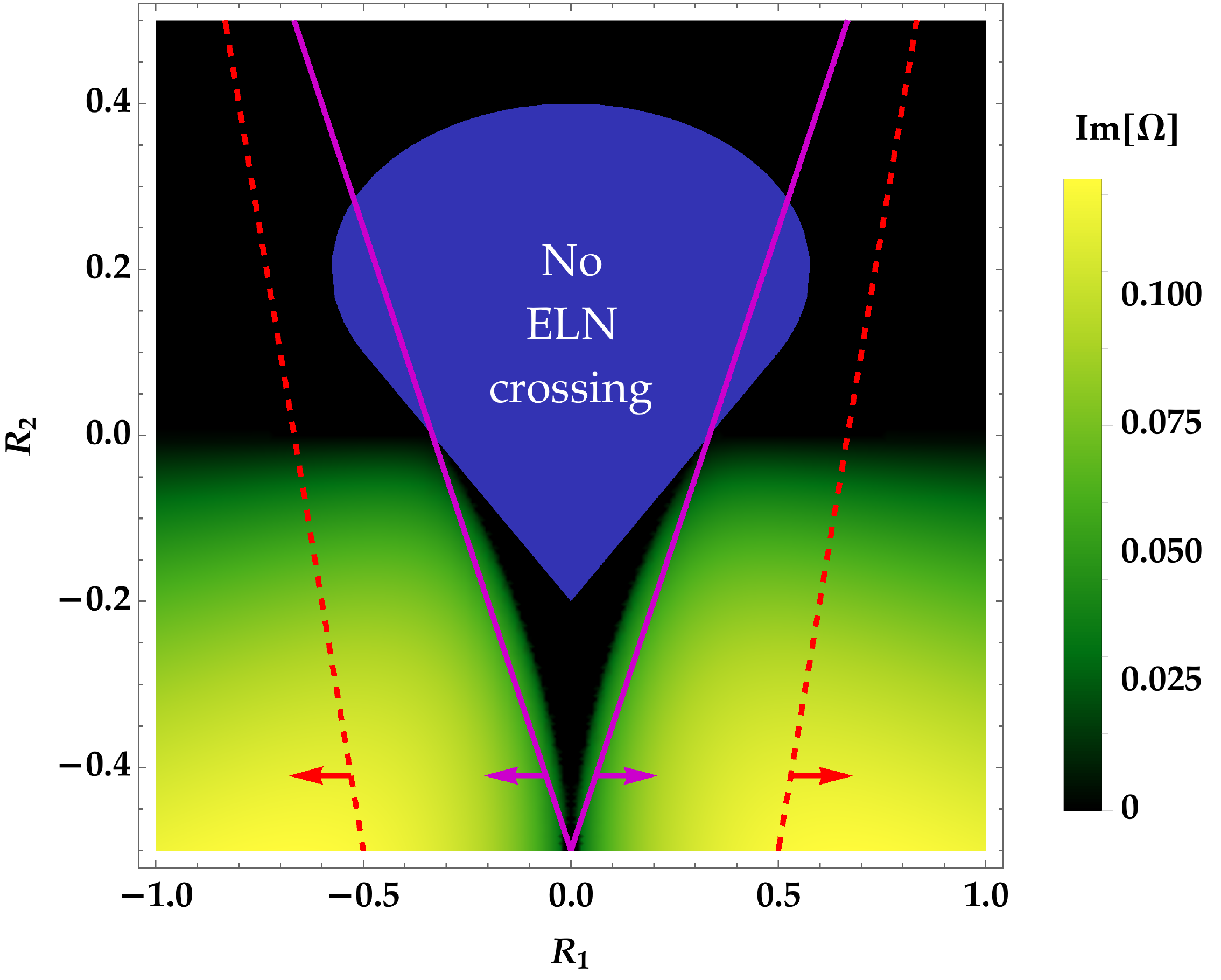}
}
\end{subfigure}
\begin{subfigure}{
\centering
\includegraphics[width=.315\textwidth]{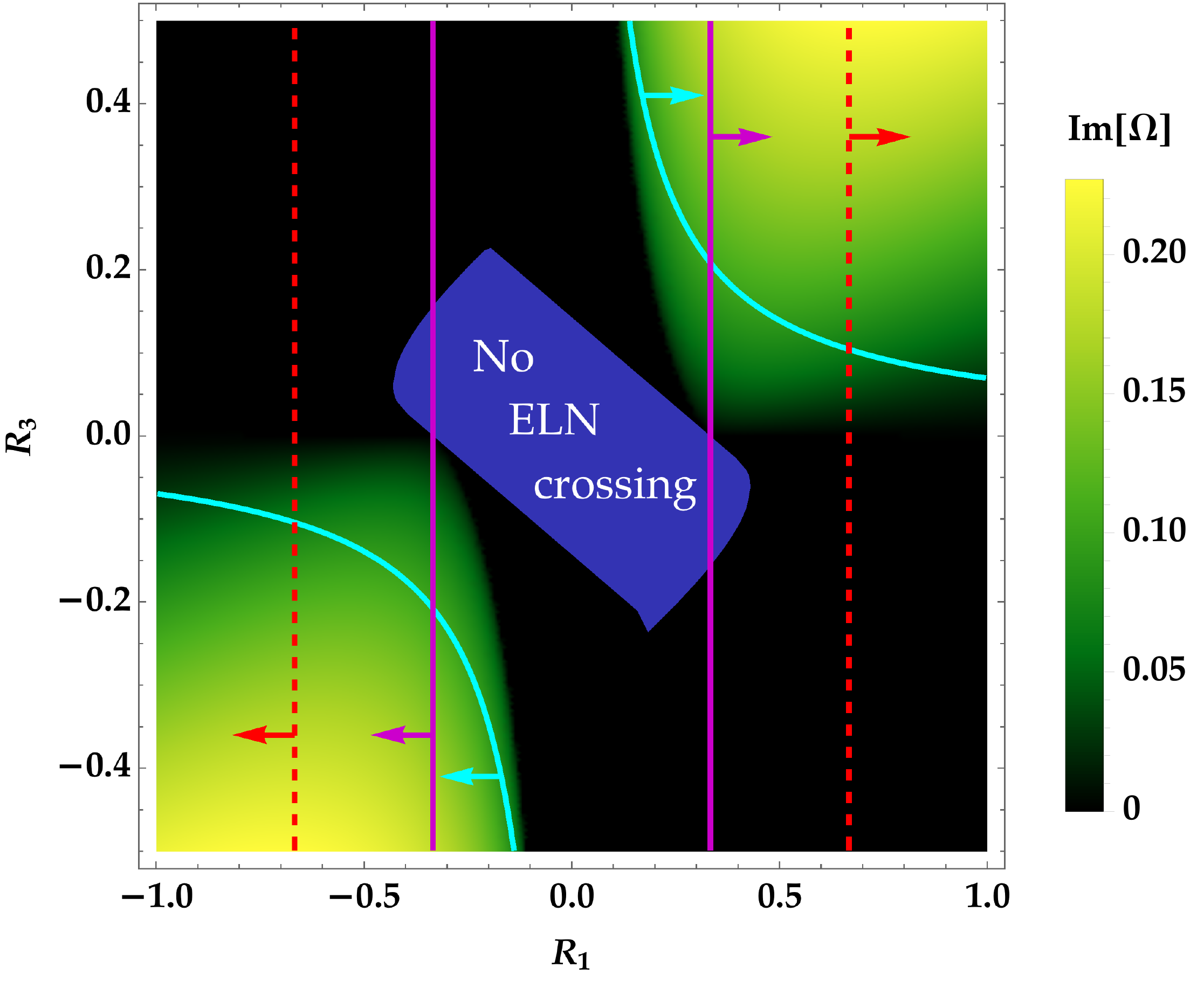}
}
\end{subfigure}
\begin{subfigure}{
\centering
\includegraphics[width=.30\textwidth]{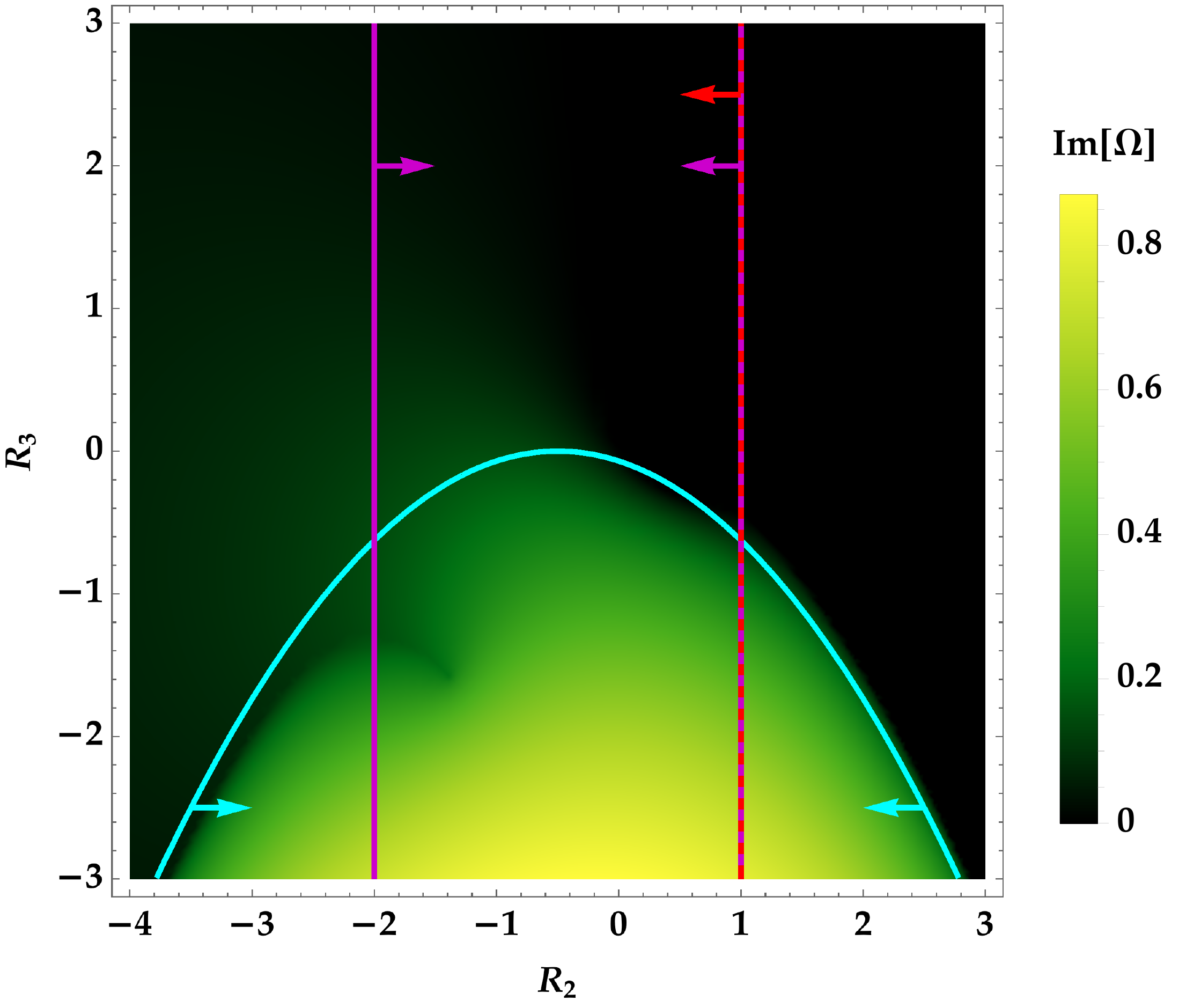}
}
\end{subfigure}
\caption{Regions of instability. Each point represents a family of angular distributions (Eq.~\eqref{ratiosdef}). \textit{Left}: The $(R_1, R_2)$ parameter space, with $R_{l\geq3} = 0$. \textit{Center}: $(R_1,R_3)$, with $R_2 = R_{l\geq4} = 0$. \textit{Right}: $(R_2,R_3)$, with $R_1 = -1$ and $R_{l\geq4} = 0$. The color map shows the instability growth rate obtained from the linear analysis [Eq.~\eqref{dispersion}] in units of $\sqrt{2} G_F (n_{\nu_e} - n_{\bar{\nu}_e})$; the blue region indicates parameters for which no zero-crossing occurs in the electron lepton number carried by neutrinos; and the magenta, cyan, and red curves border the unstable regions according to Eqs.~\eqref{vtilde}, \eqref{southernmost}, and \eqref{zeromode}, respectively. Arrows point \textit{into} the unstable regions. For reference, the Fig.~\ref{angles} angular distributions at 33 km have $R_1 = -0.35~(-1.11)$, $R_2 = -0.02~(-0.05)$, and $R_3 = 0~(0)$ for $\alpha = 0.97~(0.99)$. The angular distributions at 70 km have $R_1 = -0.17~(-0.87)$, $R_2 = -0.24~(-0.71)$, and $R_3 = -0.12~(-0.34)$ for $\alpha = 0.90~(0.95)$. }  
\label{stability}
\end{figure*}

Conducting a linear stability analysis in terms of angular moments is revealing as well. Following the usual procedure \cite{banerjee2011}, we take the flavor coherence to be of the collective form $S_{E,v} = Q_{E,v} \exp (- i \Omega t )$ and search for growing solutions ($\textrm{Im}~\Omega > 0$) to the dispersion relation
\begin{equation}
( 1 + I_0 ) (1 - I_2) + I_1^2 = 0, \label{dispersion}
\end{equation}
where
\begin{gather}
I_j = \sqrt{2} G_F \left( n_{\nu_e} - n_{\bar{\nu}_e} \right) \sum_{l=0}^\infty \left( l + \frac{1}{2} \right) R_l I_{j,l}, \notag \\
I_{j,l} = \int_{-1}^1 dv \frac{v^j L_l (v)}{\Omega - \sqrt{2} G_F \left( n_{\nu_e} - n_{\bar{\nu}_e} \right) \left( 1 - R_1 v \right)}.
\end{gather}
We continue to set $\lambda = \omega = 0$, and we assume that $n_{\nu_x} = n_{\bar{\nu}_x}$. In these expressions $L_l$ is the $l$\textsuperscript{th} Legendre polynomial and $R_l$ is the ratio of the $l$\textsuperscript{th} Legendre moment of the $\nu$ELN to the total $\nu$ELN (\textit{i.e.}, $R_l$ is the same parameter that appears in Eq.~\eqref{vtilde}):
\begin{equation}
R_l = \frac{\left( n_{\nu_e} - n_{\bar{\nu}_e} \right)_l}{n_{\nu_e} - n_{\bar{\nu}_e}}. \label{ratiosdef}
\end{equation}
As the pendulum analysis suggests, it \textit{is} possible to have instability with $n_{\nu_e} = n_{\bar{\nu}_e}$, but for the convenience of working with dimensionless ratios whose meanings are fairly transparent, we assume that the number densities are not extremely close in value. Since $\sqrt{2} G_F ( n_{\nu_e} - n_{\bar{\nu}_e} )$ only serves to set the time scale, stability is entirely controlled by the parameters $R_{l \geq 1}$.

One virtue of assessing stability in terms of angular moments is that any $I_{j,l}$ (or the equivalent when $\omega \neq 0$) can be evaluated analytically, thereby preserving the singularity structure. The singular feature in this case is a branch cut along the real axis of the complex-$\Omega$ plane; it spans the values for which the integrand of $I_{j,l}$ diverges for some $v \in [-1, 1]$. By retaining the logarithms in Eq.~\eqref{dispersion}, one avoids the unwelcome appearance of spurious instabilities \cite{morinaga2018}. We suspect that this advantage carries over to nonlinear calculations that directly evolve the angular moments.

As for what the stability analysis reveals, we find that it qualitatively bears out the $\mathbf{D}_1$ pendulum dynamics. The primary features of Fig.~\ref{stability}, which presents the regions of instability in three different ways, are all accounted for by Eqs.~\eqref{deltaeq} and \eqref{conserved}. In brief, the main takeaway is that the system is destabilized if the $l=2$ moment of the $\nu$ELN has the \textit{opposite} sign to the $l=0$ moment (because the spin $\sigma$ is thereby diminished, up to a point) or if the $l=3$ moment has the \textit{same} sign as the $l=0$ moment (because then $\mathbf{G}' (0) \cdot \boldsymbol{\delta}' (0) > 0$ and the pendulum is initially inverted). The liminal case $R_2 = 0$ in the leftmost color map is also necessarily stable, because $\mathbf{D}_3$ never becomes nonzero: no gravity, no instability. A related observation can be made about the numerical solution of the nonlinear equations, where we have confirmed that FFC occurs when the system is truncated at $l = 3$ but disappears when the system is truncated at $l = 2$.

While a $\nu$ELN crossing is commonly believed to be a necessary condition for FFC \cite{sawyer2016, chakraborty2016c, dasgupta2017, abbar2018}, Fig.~\ref{stability} shows that it is not a sufficient one. An alternative estimate of the unstable region can be obtained by supposing that $\mathbf{D}_3$ is constant. Using conservation of energy and conservation of angular momentum along $\mathbf{D}_3$, we can solve for the southernmost deviation $\theta_{1,\textrm{max}}$ reached by an initially inverted pendulum \cite{hannestad2006}:
\begin{equation}
\cos \theta_{1,\textrm{max}} = \frac{9 \sigma^2}{D_1 D_3} - 1. \label{th1max}
\end{equation}
Solutions disappear in the stable region of parameter space. In terms of $\nu$ELN ratios, the system is unstable if
\begin{equation}
R_1 R_3 \geq \frac{5}{72} \left( 1 + 2 R_2 \right)^2. \label{southernmost}
\end{equation}
In Fig.~\ref{stability} we compare Eq.~\eqref{southernmost} to the exact results from linear stability analysis and to the $| \tilde{v} | \leq 1$ criterion [Eq.~\eqref{vtilde}].

A different stability test was recently proposed in Ref.~\cite{dasgupta2018}, one which (like $\tilde{v}$) involves only the $l \leq 2$ $\nu$ELN angular moments. To make contact with that work, we now allow for spatially inhomogeneous collective modes: $S_{E,v} = Q_{E,v} \exp (- i \Omega t + i K r)$. In the linear regime, the only change to the foregoing results is that a term $-Kv$ is added to the denominator of $I_{j,l}$. A central insight of Ref.~\cite{dasgupta2018} is that $K = \sqrt{2} G_F ( n_{\nu_e} - n_{\bar{\nu}_e} ) R_1$ cancels the other term proportional to $v$, turning a transcendental dispersion relation into a quadratic equation. In our notation, they find the instability criterion
\begin{equation}
R_1^2 > \frac{\left( 2 + R_2 \right)^2}{9}. \label{zeromode}
\end{equation}
We plot this result in Fig.~\ref{stability} as well, bearing in mind that it is being compared to the $K = 0$ mode. The comparison should therefore be interpreted with suitable caution. In our view, all of these criteria are complementary, and they are bound to have more or less diagnostic power depending on factors such as the neutrino angular distributions and the spectrum of inhomogeneities.

Continuing in the same vein, we now show that \textit{spatially} growing, steady-state fast modes have pendulum-like behavior built into their equations of motion as well. The replacement for Eq.~\eqref{hybrid} is
\begin{equation}
\dot{\mathbf{P}}_v = \mu \left( \frac{1}{v} \mathbf{D}_0 - \mathbf{D}_1 \right) \times \mathbf{P}_v,
\end{equation}
where $v \neq 0$ and the overdot now denotes a spatial derivative. (Homogeneity along the transverse directions requires that $v = 0$ trajectories exhibit no flavor transformation.) It is again possible to rotate out $\lambda$---and we have done so---provided that we work in the nearly homogeneous limit. More precisely, we ignore small-scale fluctuations and assume that the scale heights of $\lambda$ and $\mu$ are much greater than any fast oscillation length, so that the two parameters are approximately constant over the region we consider.

Dividing through by $v$ leads, after taking angular moments, to equations that each contain a derivative of a single $l$:
\begin{equation}
\dot{\mathbf{P}}_l = - \mu \mathbf{D}_1 \times \mathbf{P}_l + \mu \mathbf{D}_0 \times \sum_{l' = 0}^\infty \left( l' + \frac{1}{2} \right) c_{l l'} \mathbf{P}_{l'}, \label{hybridspace}
\end{equation}
where
\begin{equation}
c_{l l'} = \int_{-1}^{1} dv \frac{L_l (v) L_{l'} (v)}{v}.
\end{equation}
To make sure the integrals converge, we interpret them as denoting their principal values, or equivalently assert that $\mathbf{P}_v = 0$ at $v=0$. We presume that the collective modes of the system are not particularly sensitive to the flavor distribution of neutrinos traveling \textit{precisely} transverse to the symmetry axis. From the orthogonality and recursion relations of Legendre polynomials, it follows that
\begin{equation}
c_{l l'} = \begin{cases}
\frac{2}{l+1} \prod (-1) \frac{l' -2n + 1}{l' - 2n +2}   &\textrm{odd}~l' > \textrm{even}~l \\
\frac{2}{l} \prod (-1) \frac{l' + 2n}{l' + 2n - 1}   &\textrm{even}~l' < \textrm{odd}~l \\
0   &\textrm{otherwise}.
\end{cases} \label{cll}
\end{equation}
The product in both cases is from $n=1$ up to $n = ( | l - l' | - 1 ) / 2$ and is equal to $1$ if $| l - l' | = 1$.

An immediate consequence of Eqs.~\eqref{hybridspace} and \eqref{cll} is that $\mathbf{D}_1$ is constant. It is therefore possible to shift to a rotating frame in which the $-\mu \mathbf{D}_1 \times \mathbf{D}_l$ terms drop out. Letting primes denote the new frame, we introduce (or, rather, repurpose) the vectors
\begin{gather}
\mathbf{L}' = - \sum_{l'} \left( l' + \frac{1}{2} \right) c_{0 l'} \mathbf{D}'_{l'}, \notag \\
\mathbf{G}' = \sum_{l', l''} \left( l' + \frac{1}{2} \right) \left( l'' + \frac{1}{2} \right) c_{0 l'} c_{l' l''} \mathbf{D}'_{l''}, \notag \\
\boldsymbol{\delta}' = \frac{\mathbf{D}'_0}{D_0}, ~~~~ \sigma = \boldsymbol{\delta}' \cdot \mathbf{L}',
\end{gather}
Calculating $\boldsymbol{\delta}' \times \ddot{\boldsymbol{\delta}}'$, we find ourselves back at Eq.~\eqref{deltaeq}, but with $D_1$ replaced by $D_0$. Once again the pendulum's length, spin, and mechanical energy (given by Eq.~\eqref{conserved} after sending $\mathbf{D}'_1 \rightarrow \mathbf{D}'_0$) are all conserved. Besides this replacement, there is another fundamental difference between the temporal and spatial flavor development: Eq.~\eqref{cll} tells us that $\mathbf{L}'$ is a superposition of \textit{all} odd moments, whereas $\mathbf{G}'$ is a superposition of \textit{all} even moments. Inhomogeneity brings a host of complications with it, and so we leave for future work the task of exploring numerically how the pendulum-like tendency manifests in spatially evolving collective modes.

The aim of this study has been to extract analytic insights into FFC from the nonlinear equations of motion. The central finding is that the angular-moment equations exhibit a certain pendulum-like structure in the two limits which are most analytically tractable (\textit{viz.}, when the neutrino density is high, the matter background is homogeneous, and the neutrino flavor field is \textit{either homogeneous or stationary}). In general, of course, a flavor field develops both spatially \textit{and} temporally. More work must be done to understand what our finding implies for the full PDE problem.

The analysis presented here opens new paths toward understanding collective oscillations and incorporating their effects into frontline supernova simulations.

\begin{acknowledgments}
We are grateful to David Radice for making the \textsc{Fornax} data accessible, and to Vincenzo Cirigliano, Pat Diamond, Mark Paris, and David Vartanyan for enlightening discussions. HN acknowledges Taiki Morinaga, Shoichi Yamada and Sherwood Richers for valuable comments and discussions. LJ and GMF acknowledge support from NSF Grant Nos. PHY-1614864 and PHY-1914242, from the Department of Energy Scientific Discovery through Advanced Computing (SciDAC-4) grant register No. SN60152 (award number de-sc0018297), and from the NSF N3AS Hub Grant No. PHY-1630782 and Heising-Simons Foundation Grant No. 2017-22. AB acknowledges support from the U.S. Department of Energy Office of Science and the Office of Advanced Scientific Computing Research via the Scientific Discovery through Advanced Computing (SciDAC4) program and Grant DE-SC0018297 (subaward 00009650). In addition, he acknowledges support from the U.S. NSF under Grants AST-1714267 and PHY-1144374.
\end{acknowledgments}

\bibliography{all_papers}

\end{document}